\documentclass[10pt]{article}
\usepackage[cp1251]{inputenc}
\usepackage[english]{babel}
\usepackage{graphicx}
\usepackage{amssymb}
\usepackage{amsmath}
\usepackage{amsthm}
\usepackage{amsfonts}
\usepackage{amssymb}
\title{{\bf \Large  Cosmological Models with a Varying  $\Lambda$ -Term \\ in Lyra's Geometry}\\
{\normalsize ~~{\bf V.\,K. Shchigolev}\thanks{E-mail:
vkshch@yahoo.com}}\\
{\small {\it Ulyanovsk State University, 42 L. Tolstoy Str.,
Ulyanovsk 432000, Russia}}\\
\vspace{2mm}
\small \begin{quote}{\bf Abstract} --  Cosmological models in  Lyra's geometry are constructed and investigated with the assumption of a minimal interaction  of  matter  with  the  displacement vector field and the dynamical  $ \Lambda $  term.  Exact solutions of the model equations are obtained  for the different equations of state of the matter, that fills the universe, and for the certain assumptions on the decaying law for $\Lambda$.
 \\
\vspace{2,5mm}
{\bf PACS numbers}: 98.80.-k; 98.80.Jk; 04.20.Jb.\\
{\bf Key words}: cosmological models, Lyra's geometry, cosmological term, accelerated expansion.\\
\end{quote}}
\date{}

\begin{document}

\maketitle
\vspace{-2.5cm}
\section{Introduction}

~~~~The study of the cosmological models with a varying  $\Lambda$ -term has been essentially  activated by the efforts of several researchers from as long as a cosmological constant problem gained a real confirmation of decaying $\Lambda (t)$. There are a large number of observations for the determination of Einstein's cosmological constant, $\Lambda$, or some matter ingredient of the Universe which is slowly changing with time and, thus, acting like $\Lambda $.
Moreover, recent observations of supernovae of type Ia \cite{Riess, Perlmutter}, Cosmic Microwave Background Radiation \cite{Spergel, Komatsu}, Baryon Acoustic Oscillations in galaxy surveys \cite{Blake, Seo} etc., are  of evidences in favor of a non-zero cosmological 'constant' \, with relative energy density $\Omega_{\Lambda} = \Lambda/3H_0^2 \approx 0.6 - 0.7 $. This value could not remain constant during long-term observation. The recent studies of nonlocal effects, wormholes, inflationary mechanisms of cosmological perturbations  are the evidence in favor of decreasing with time of  effective cosmological term.
Cosmological models with a time-varying $\Lambda(t)$ with different decay laws were proposed by several researchers during the last two decades (see, e.g.  \cite{Chen} -\cite{Carneiro}).

However, the formal introduction of the dynamical $ \Lambda $ - term into Einstein's equation of General Relativity (GR) leads to the violation of the energy conservation law of matter. This follows directly  from the Einstein's field equation due to the Bianchi identity for the Riemann curvature tensor, if the cosmological term is constant or zero. Previously, several authors have investigated cosmological models in Lyra's geometry \cite {Lyra}, which is a generalization of Riemannian geometry by introducing a gauge function which removes the non-integrability of the length of the parallel transfer characteristic of Weyl's theory (see, eg, \cite{Singh} and references therein). It was noted that cosmology based on Lyra's manifold with constant gauge vector is similar to the C-field theory of Hoyle-Narlikar \cite{Hoyle}, or contains a vacuum field, which together with the gauge vector field can be considered as a cosmological term. A scalar-field cosmological model in Lyra's geometry is investigated in \cite {Shchigolev}, where it was also noted that the dynamical displacement field, which is free from interaction with matter and does not violate the energy conservation law of matter, can only serve as a stiff perfect fluid. Thus, we can assume that the simultaneous consideration of the dynamical  $\Lambda$ - term and displacement field is able not only to  prevent a violation of the energy conservation law but also to enrich the theory. In our view, this approach could  lead to the significant variations in evolution of the standard cosmological model.

The purpose of this paper is to construct a cosmological model in Lyra's geometry, provided to the minimal interaction of matter with the displacement vector field and dynamical $\Lambda$ - term.
We find some exact solutions of the model equations for the different states of matter that fills the universe, and certain assumptions regarding the evolution of the cosmological term. In addition, we analyze the possibility of the accelerated cosmological expansion on the basis of solutions obtained here.

\section{A Model with Time Evolving Cosmological Term in Lyra's Geometry}

~~~~The Einstein's field equations with a cosmological $\Lambda$ - term in Lyra's geometry, as proposed in \cite{Sen} in normal gauge, may be written as
\begin{equation}\label{1}
R_{ik}- \frac{1}{2} g_{ik} R - \Lambda g_{ik} +  \frac{3}{2}\phi_i \phi_k - \frac{3}{4}g_{ik}\phi^j \phi_j = T_{ik},
\end{equation}
where $\phi_i$ is the displacement vector, the gravitational constant is $8\pi G=1$, and other symbols have their usual meanings in the Riemannian geometry.
The energy-momentum tensor (EMT) of matter $T_{ik}$ can be derived in a usual manner from the Lagrangian of matter. Considering the matter as some effective perfect fluid, we can write:
\begin{equation}\label{2}
T_{ik}= (\rho_m +p_m)u_i u_k -p_m\, g_{ik},
\end{equation}
where  $u_i = (1,0,0,0)$ is the 4-velocity of the co-moving observer, satisfying $u_i u^i = 1$.
Then let us represent $\phi_i$ as a time-like vector field of displacement,
\begin{equation}\label{3}
\phi_i = \left(\frac{2}{\sqrt{3}}\,\beta,0,0,0\right),
\end{equation}
where $\beta = \beta(t)$ is a function of time alone, and the factor $2/\sqrt{3}$ is introduced to simplify the subsequent equations. The metric of a Friedmann-Robertson-Walker (FRW) space-time can be written as follows: $ds^2 = d t^2- a^2 (t)(d r^2+\xi^2 (r)d \Omega^2)$, where $a(t)$ is a scale factor of the Universe, $\xi(r)=\sin r,r,\sinh r$ in accordance with the  sign of the spatial  curvature $k=+1,0,-1$.
Given this metric and Eqs. (\ref{2}), (\ref{3}), the field equation (\ref{1}) can be reduced to the following set of equations:
\begin{equation}\label{4}
3H^2 + \frac{3 k}{a^2} - \beta^2 = \rho_m +\Lambda,\,\,\,
2 \dot H + 3H^2 + \frac{k}{a^2} + \beta^2 = -  p_m+\Lambda,
\end{equation}
where $H = \dot a/a $ is the Hubble parameter, and an overdot denotes  differentiation with respect to time $t$.

As a consequence of Eq. (\ref{4}),  the continuity equation for the effective matter can be written as:
\begin{equation}\label{5}
\dot \rho_m + \dot \Lambda + 2 \beta \dot \beta + 3 H \Big[\rho_m + p_m + 2\beta^2 \Big]=0.
\end{equation}
One of the most important quantity to describe the features of dark energy models
is the equation of state (EoS)
It is known that the most important quantity to describe the possibility of an accelerated mode of expansion is the so-called deceleration parameter $q$, defined as
\begin{equation}\label{6}
q=- \frac{a \ddot a}{\dot a^2}=-1-\frac{\dot H}{H^2}.
\end{equation}

To proceed further in studying of our model, it is necessary to determine the type of dependence of the cosmological term (or the displacement vector field) on time, or the type of interaction between the geometric fields and matter.
For simplicity, we consider a spatially flat FRW cosmology with $k=0$.
We can write the basic equations of the model (\ref{4}) as follows:
\begin{equation}\label{7}
3H^2  = \rho_m +\Lambda + \beta^2,~~
2 \dot H  = - (\rho_m + p_m + 2  \beta^2),
\end{equation}
and the continuity equation (\ref {5}) is in the same form. The latter follows from the set of equations (\ref {7}).

The main assumption on our model consists of the minimal coupling of the matter with the displacement vector field and the cosmological term on Lyra's manifold. That means that there is no the direct interaction between them, that is this interaction is realized only through the gravitational field. It allows us to avoid the violation of the energy-momentum conservation of matter in the framework of minimal demands on  the  behavior of displacement vector. Indeed, due to the covariant equation of  the energy-momentum conservation $T^k_{i\,; k} = 0$ and the identity $G^k_{i\,; k} = 0$ for the Einstein tensor, the field equation (\ref{1}) leads to the following equation for the cosmological term and the displacement field:
\begin{equation}\label{8}
 \Lambda_{;i} -  \frac{3}{2}\Big(\phi_i \phi^k - \frac{1}{2}\delta_i^k\phi^j \phi_j\Big)_{;k}=0.
\end{equation}
In the absence of cosmological term (or in the case of its constancy), when $\Lambda_{;\, i}\equiv 0$, Eq. (\ref{8}) leads to the interpretation of the displacement field as an analog of the so called stiff fluid with the EoS $w_{sf} = $1. The assumption of the non-vanishing (and varying with time) cosmological term can significantly extend the capabilities of such a model in describing of the real processes taking place in the Universe.

Here, the minimal coupling means that  the energy conservation law for matter is valid regardless of the presence of $\Lambda (t)$ and $\beta(t)$:
\begin{equation}
\dot \rho_m + 3 H (\rho_m + p_m )=0.\label {9}
\end{equation}
Thus, the remaining part of Eq. (\ref{5}), i.e.  Eq. (\ref{8}), has the form:
\begin{equation}\label{10}
\dot \Lambda + 2 \beta \dot \beta + 6 H \beta^2 =0.
\end{equation}

The set of equations (\ref{7}), (\ref{9}), (\ref {10}), which determines the dynamics of our model, must be supplemented by some conditions. In our view, the most realistic conditions are: first,  the EoS of matter $p_m = w_m \rho_m$ and, second, the law of evolution of the cosmological term $\Lambda(t)$ which should correspond to the observational data. Of course, this case is not the only one in  searching of exact solutions in the framework of such a model. Nevertheless,  in the present paper we study our model just with  this assumption, as  we have no any possible dependence $\beta(t)$,  which could be proved by the direct observations.

We assume that the matter content of the Universe can be considered as a barotropic perfect fluid with a constant equation of state (EoS): $-1 \leq w_m \leq 1$. Then Eq. (\ref{9}) can be easily integrated which yields
\begin{equation}\label{11}
\rho_m = \rho_0 a^{\displaystyle -3(1+w_m)},
\end{equation}
where $\rho_0$ is a constant of integration.
Due to (\ref{11}), the equations (\ref{7}) can be written as follows:
\begin{eqnarray}
3H^2  &=& \rho_0 a^{\displaystyle -3(1+w_m)} +\Lambda + \beta^2, \label{12}\\
2 \dot H  &=& - (1+w_m)\rho_0 a^{\displaystyle -3(1+w_m)} - 2  \beta^2, \label{13}
\end{eqnarray}
and Eq. (\ref{10}), as a differential consequence of Eqs. (\ref{12}), (\ref{13}), is not changed, that  can be easily verified.

Thus, the problem is reduced to solving the set of equations (\ref{12}), (\ref{13}) either for $ H (t) $ and $\beta(t)$ with some given function $\Lambda(t)$, or for  $H(t)$ and $\Lambda(t)$ with a given function $\beta(t)$. Since no information about the possible dependence of $\beta(t)$ is available, we consider the phenomenological models with some given functions $\Lambda(t)$ that have an observational ground and are widely discussed in the literature (see, e.g. \cite{Sahni}), namely: $\displaystyle (i) ~ \Lambda(t) = \frac{\alpha}{t^2}$, and $\displaystyle  (ii) ~ \Lambda(t) = \frac{\alpha}{t} H(t)$.  Besides, two cases, $w_m = -1$ and $w_m\ne-1$,  should be considered separately.

\begin{figure}[t]
\includegraphics[width=0.47\textwidth]{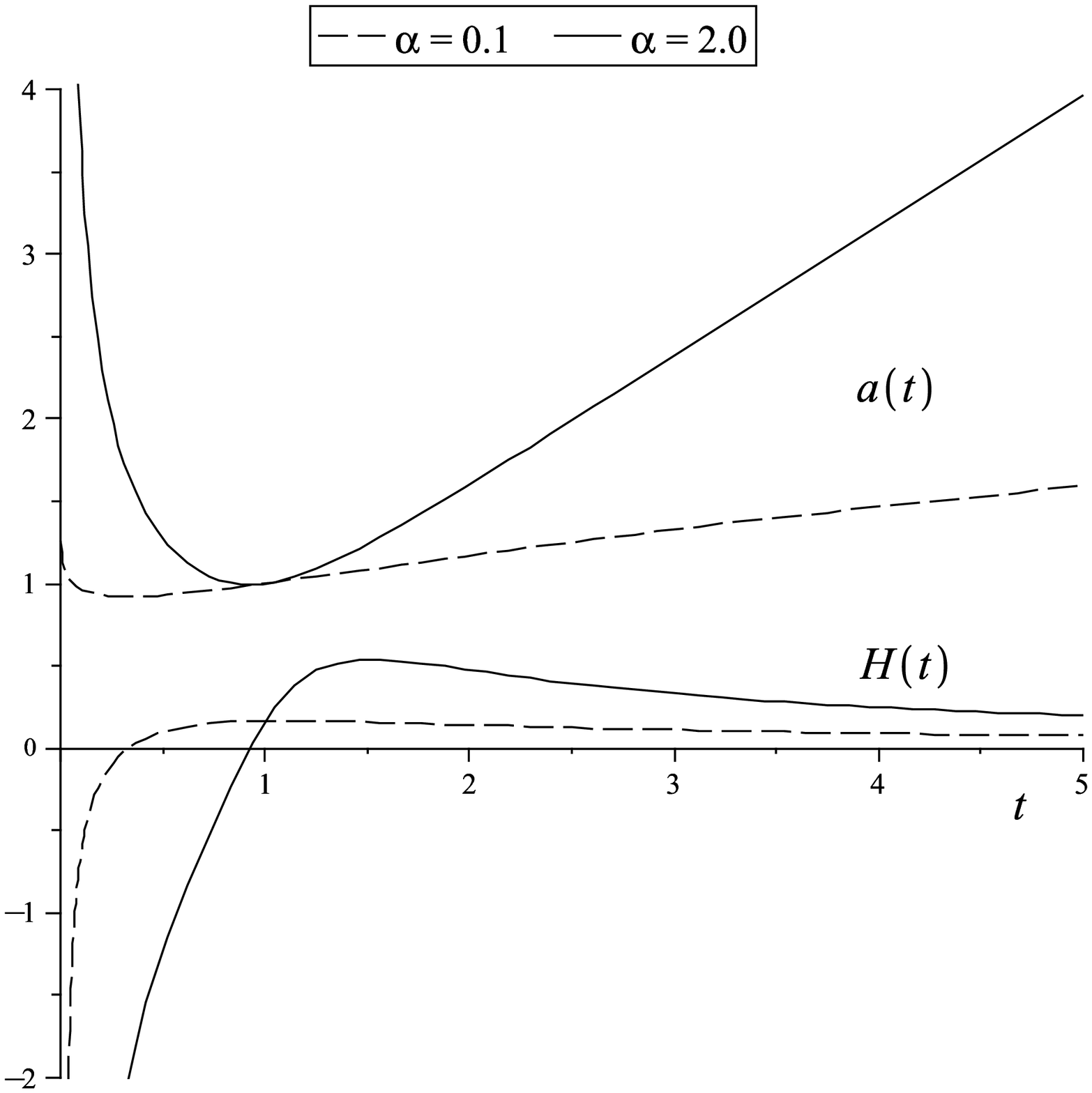} \hfill
\includegraphics[width=0.47\textwidth]{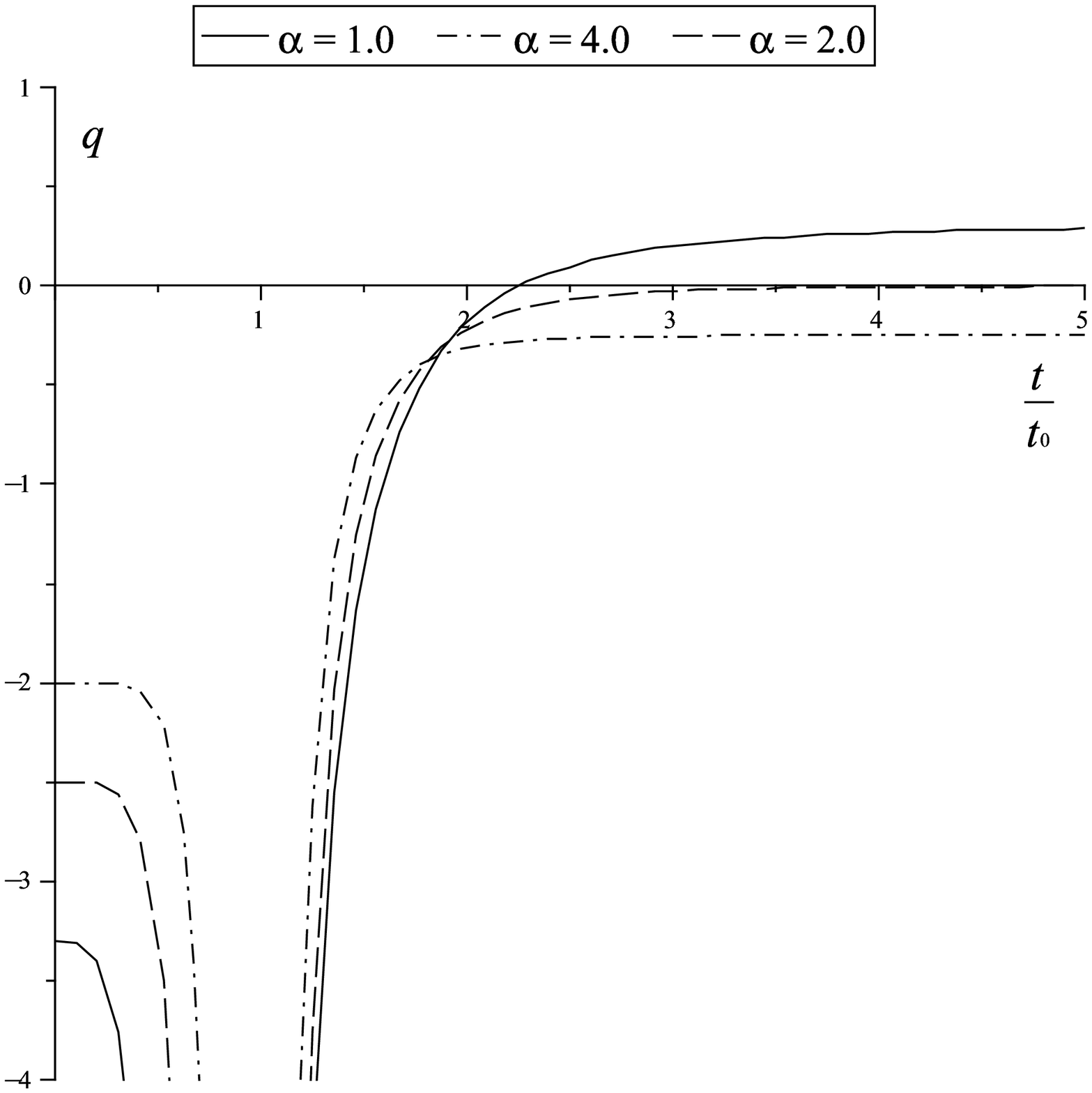}
\parbox[t]{0.47\textwidth}{\caption{Plots of $a(t)$ and $H(t)$ vs time for the $w_m = -1$ model in the case  $\Lambda_{eff}=\alpha/t^2$. Here $t_0 = 1$.}\label{fig1}} \hfill
\parbox[t]{0.47\textwidth}{\caption{The plot of $q(t)$ vs time for the $\Lambda_{eff}=\alpha/t^2$ model.}\label{fig2}}
\end{figure}

\section{Models with the EoS of quasi-vacuum: $w_m=-1$}
~~~~In this case, Eq. (\ref{11}) leads to the usual property of the quasi-vacuum state: $p_m=-\rho_m=-\rho_{\mbox{v}} = constant$. Then, Eqs. (\ref{12}) and (\ref{13}) can be rewritten as
\begin{equation}
 \dot H + 3H^2  = \rho_{\mbox{v}} +\Lambda, \, \, \,\, \beta^2 = - \dot H. \label{14}
\end{equation}
It is easy to see that the constant energy density  of the quasi-vacuum  $\rho_0$ can be added to the cosmological term. By introducing notation  $\Lambda_{eff} = \rho_{\mbox{v}} + \Lambda$,  we can obtain from Eq. (\ref{14}):
$\dot H + 3H^2 = \Lambda_ {eff}$.
We consider two cases depending on $\Lambda_ {eff}(t)$, mentioned above.

As a result of solving Eq. (\ref{14}) with $\Lambda_ {eff} = \alpha /t^2$ for the case $(i)$, we obtain the following expressions:
\begin{eqnarray}
H(t) & =& \frac{1}{6 t}\left\{1+ n_{\alpha}  \,\tanh\Big[\frac{n_{\alpha}}{2}\ln(t/t_0)\Big]\right\} ,\,\,n_{\alpha}=\sqrt{1+12\alpha},\label{15}\\
a(t) &=& a_0 \Big(\frac{t}{t_0}\Big)^{1/6} \cosh^{1/3}\Big[\frac{n_{\alpha}}{2}\ln\Big(\frac{t}{t_0}\Big)\Big],\label{16}\\
\beta^2(t) &=& \frac{1}{12 t^2}\left\{\Big(1+ n_{\alpha} \,\tanh\Big[\frac{n_{\alpha}}{2}\ln(t/t_0)\Big]\Big)^2+1-n_{\alpha}^2\right\}.\label{17}
\end{eqnarray}
The plots of  $a(t)$ and $H(t)$ for this solution with two different values of the coupling constant $\alpha$ are shown in Fig. 1.

It is easy to find that the deceleration parameter (\ref{6}) is equal to
\begin{equation}\label{18}
q(t)=2-3\frac{n_{\alpha}^2-1}{\left\{1+ n_{\alpha}  \,\tanh\Big[\displaystyle \frac{ n_{\alpha}}{2}\ln(t/t_0)\Big]\right\}^2}
\end{equation}
due to Eq. (\ref{15}).
The plot of $q(t)$ for various values of the coupling constant $\alpha$ is shown in Fig. 2. One can see that for each value of $\alpha$ there exists a finite positive value $t_{cr}$, at which the deceleration parameter is the negative infinity. This $t_{cr}$ can be found from the equality of the denominator in Eq. (\ref{18}) to zero.
Besides, from Eq. (\ref{17}) it follows that the displacement field becomes real only since a certain instant $t_i$, which can be found from equation $\beta = 0$. Thus, we have
\begin{equation}\label{19}
t_{cr} = t_0 \left(\frac{n_{\alpha}-1}{n_{\alpha}+1}\right)^{\displaystyle\frac{1}{n_{\alpha}}}\,,t_i = t_0 \left(\frac{n_{\alpha}-1+\sqrt{n_{\alpha}^2-1}}{n_{\alpha}+1-\sqrt{n_{\alpha}^2-1}}\right)^{\displaystyle\frac{1}{n_{\alpha}}}.
\end{equation}
One can see that $t_i = 0$ for $n_ {\alpha} = 1$, i.e. for $\alpha = 0 $.
The comparison of formulas in Eq. (\ref{19}) leads to the conclusion that, the inequality  $t_i>t_ {cr}$ is valid, except the trivial case $\alpha = 0$. Moreover, for $0 <\alpha \le 1/12$  we have $t_i\le t_0$, and  we have always $t_i>t_0$ for $\alpha>1/12$.
These estimates suggest that the nature of this model becomes realistic when $ t \ geq t_i $,  that is when there are no any problems with  the  displacement field to be real, and with the infinite acceleration of expansion.
\begin{figure}[t]
\includegraphics[width=0.47\textwidth]{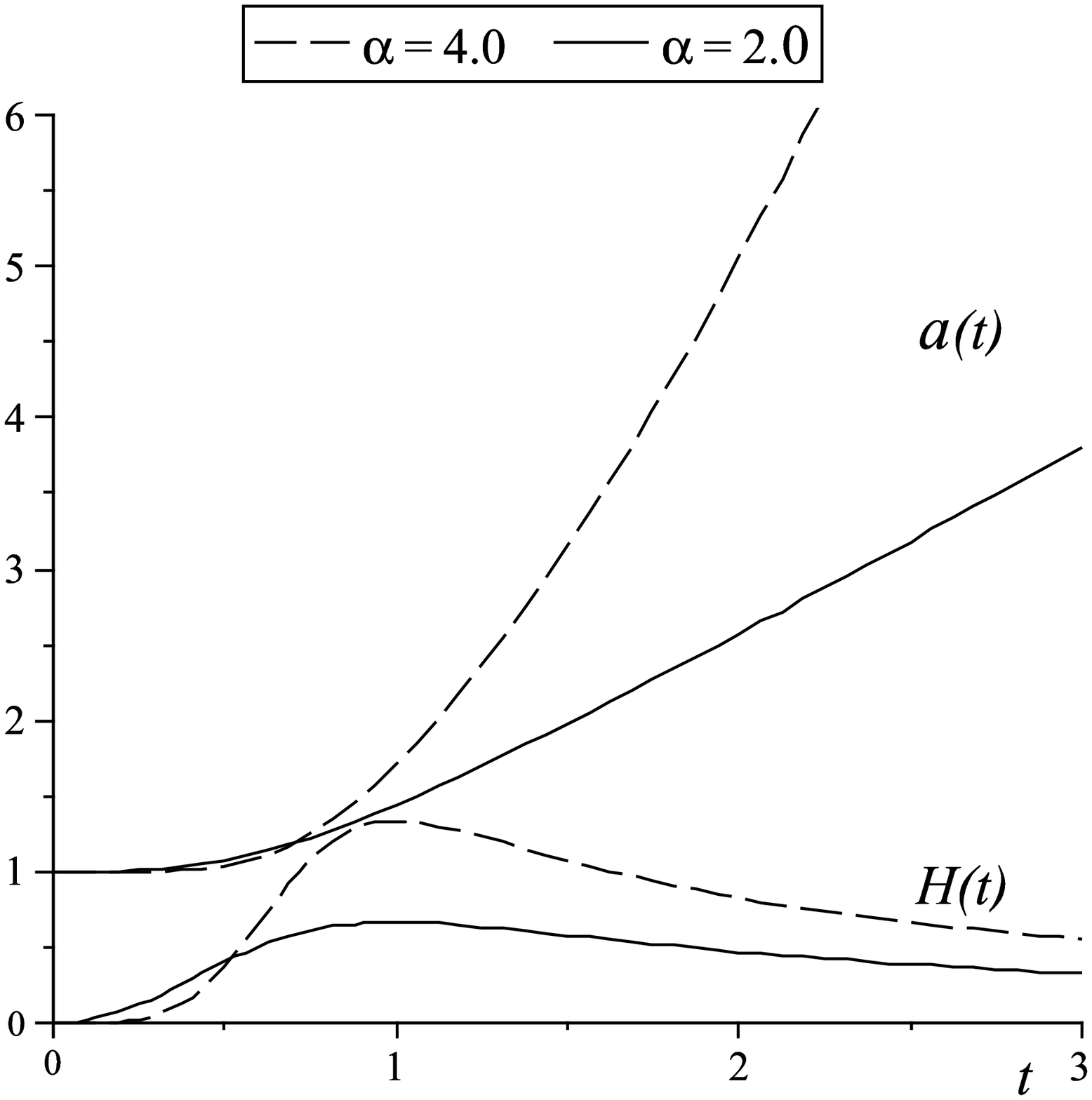} \hfill
\includegraphics[width=0.47\textwidth]{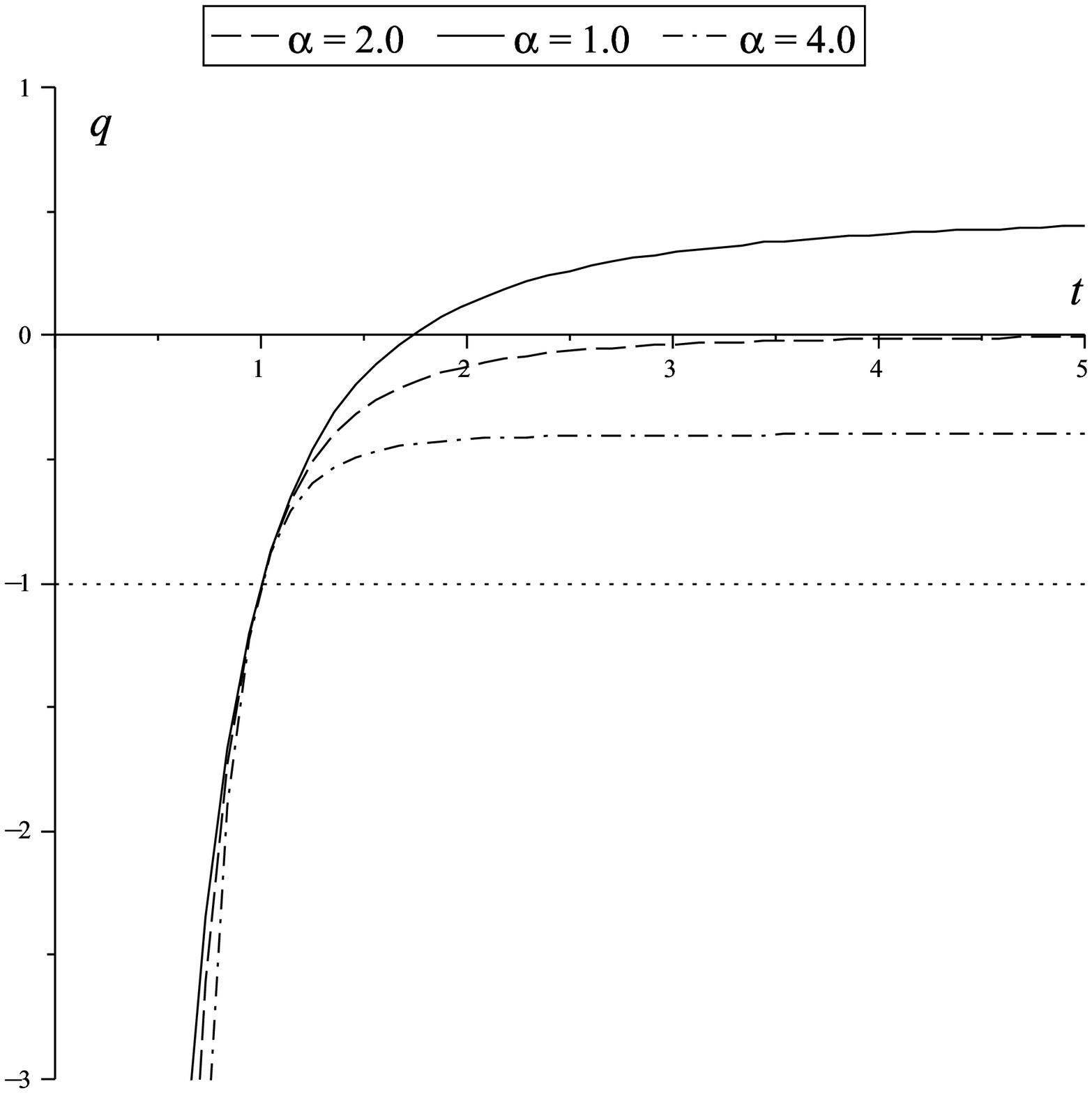}
\parbox[t]{0.47\textwidth}{\caption{Plots of $a(t)$ and $H(t)$ for the $\Lambda_{eff}=(\alpha/t)H$ model with $t_i = 1$.}\label{fig3}} \hfill
\parbox[t]{0.47\textwidth}{\caption{The plot of $q$ vs time in the $\Lambda_{eff}=(\alpha/t)H$ model for some values of $\alpha$ and $t_i = 1$.}\label{fig4}}
\end{figure}

For the case $(ii)$, we get from Eq. (\ref{14}) the following equation for the Hubble parameter:
\begin{equation}
 \dot H + 3H^2  = \frac{\alpha}{t}H, \label{20}
\end{equation}
whose solution can be easily found as follows:
\begin{equation}
\label{21} H(t) = \frac{\alpha(1+\alpha)}{3 t_i} \frac{(t/t_i)^{\alpha}}{1+\alpha(t/t_i)^{\alpha+1}},
\end{equation}
where $t_i$ is a constant of integration. Then integrating Eq. (\ref{21}) for $a(t)$, we find the scale factor in the following form:
\begin{equation}\label{22}
a(t) = a_0 \Big[1+\alpha(t/t_i)^{\alpha+1}\Big]^{1/3},
\end{equation}
where $a_0$ is a constant. Taking into account the second equation in (\ref{14}) and Eq. (\ref{21}), the following expression for the displacement field  can be found:
\begin{equation}\label{23}
\beta^2(t) = \frac{\alpha(1+\alpha)}{3 t_i^2}\frac{ (t/t_i)^{\alpha+1}-1}{[1+\alpha (t/t_i)^{\alpha+1}]^2}(t/t_i)^{\alpha-1}.
\end{equation}
From this function $\beta^2(t)$, it can be seen that $\beta^2$ changes its sign at the instant $t_i>0$. In view of the Hubble parameter (\ref{21}), we can obtain by simple mathematical manipulation that the deceleration parameter (\ref{6}) is as follows
\begin{equation}\label{24}
q(t) = -1+\frac{3}{1+\alpha}\Big[1-\Big(\frac{t_i}{t}\Big)^{\alpha+1}\Big].
\end{equation}
The plots of the scale factor $a(t)$ and the Hubble parameter $H(t)$ in this case is shown in Fig. 3. The plot of $q(t)$ versus time is shown in Fig. 4.

\section{Models with the EoS of matter $w_m\ne-1$}

~~~~Multiplying Eq. (\ref{12}) by $(1 + w_m)\ne 0$ and adding the result to Eq. (\ref{13}), we obtain the first equation  of the set of independent equations in the following form:
\begin{equation}\label{25}
2 \dot H + 3 (1+w_m) H^2 = (1+w_m)\Lambda -(1- w_m) \beta^2.
\end{equation}
We can take Eq. (\ref{10}) as the second independent equation.

For this case,  we replace the assumptions on  $\Lambda(t)$ considered above by assuming that the cosmological term  is proportional to the invariant of  the displacement vector field. Hence,  in this section  we  are supporting the idea expressed earlier about the possible role of the displacement field as an effective $\Lambda$ - term in the Einstein equation: $\Lambda = \gamma \beta^2$, where $\gamma$ is an arbitrary positive constant. Then Eq. (\ref{10}) can be integrated by introducing a new variable,
\begin{equation}\label{26}
x = \ln \Big[a(t)/a(t_{in})\Big] \,\,\,\Rightarrow\,\,\, H(t)=\frac{dx}{dt},
\end{equation}
which coincides with the number of e-folds and  means expansion of the universe  in $e^x$ times at present versus the scale factor $a_i = a(t_{in})$ at the initial time $t_{in}$. As a result, we have that
\begin{equation}\label{27}
\Lambda (x) = \Lambda_0\,\, e^{\displaystyle -\frac{6}{1+\gamma}x},\,\,\beta^2 (x) = \frac{\Lambda_0}{\gamma}\,\, e^{\displaystyle -\frac{6}{1+\gamma}x},
\end{equation}
where $\Lambda_0$ is a constant of integration. Note that it follows  from Eq. (\ref{27}) due to Eq. (\ref{26}) that the following law is realized: $\Lambda(t) \propto a(t)^{-b}$, where $b>0$. This decay law for the variation of the cosmological term was investigated earlier, for example, in \cite{Chen}, \cite{Hova}.

Substituting Eq. (\ref{27}) into Eq. (\ref{25}),  we obtain the following equation:
\begin{equation}\label{28}
 \frac{d H^2}{d x} + 3 (1+w_m) H^2 = \Lambda_0 [(1+w_m) -(1- w_m) \gamma^{-1}] e^{\displaystyle -\frac{6}{1+\gamma}x}.
\end{equation}
The general solution of this equation can be written as
\begin{equation}\label{29}
H^2= \Big(\frac{dx}{dt}\Big)^2=\frac{A}{B-b}\, e^{\displaystyle -b x} + C e^{\displaystyle -Bx},
\end{equation}
where
\begin{eqnarray}
A &=& \Lambda_0\, [(1+w_m) -(1- w_m) \gamma^{-1}] = \,2\Lambda_0\,\frac{B-b}{6-b},\nonumber\\
B &=& 3 (1+w_m),~~~~\,\, \,b = \frac{6}{1+\gamma};~~~\,~\,\,\, B,\,b\in (0;6),\label{30}
\end{eqnarray}
and $C$ is a constant of integration. In this case, as seen from Eq. (\ref{29}), we assume that $B\ne b$, i.e. $\gamma \ne \gamma_0 = (1-w_m)/(1 + w_m)$. Otherwise, i.e. for $\gamma = \gamma_0 \Leftrightarrow A = 0$, the general solution of Eq. (\ref {28}) can be written as:
\begin{equation}\label{31}
H^2= \Big(\frac{dx}{dt}\Big)^2=C e^{\displaystyle -3(1+w_m)x},
\end{equation}
where $C>0$. Taking into account the definition (\ref{26}), it is easy to  verify that the solution of Eq. (\ref{31}) reproduces the standard solution for FRW model: $a(t) \propto t ^ {2 / 3 (1 + w_m)}$.
\begin{figure}[t]
\centering
\includegraphics[width=90mm,height=6cm]{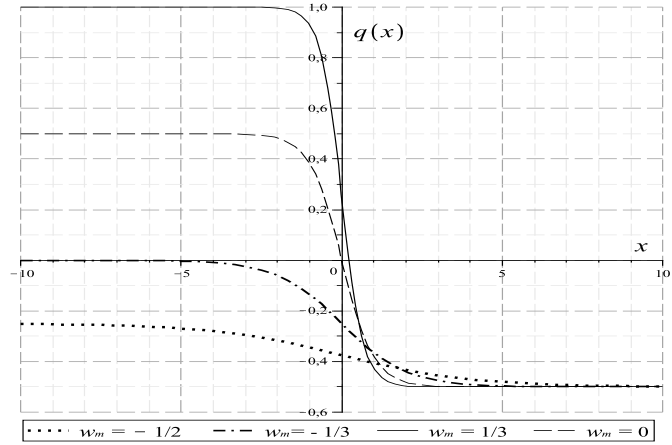}\\
\caption{Evolution of the deceleration parameter in the $\Lambda = \gamma \beta^2$ model for the different  EoS of matter $w_m\ne -1$ with $\gamma=5$ and $C_0=1$.}
\label{fig5}
\end{figure}

In the case of $\gamma \ne \gamma_0$, the general solution of Eq. (\ref{29}) can be written only in an implicit form as
\begin{equation}\label{32}
\int \limits^{\displaystyle a(t)/a_i} \frac{\xi^{\frac{B}{2}-1}\,d\,\xi}{\sqrt{C_0+\xi^{B-b}}} =\pm \sqrt{\frac{\Lambda_0}{3}\left(1+\frac{1}{\gamma}\right)}\,\,t+C_1,
\end{equation}
where $C_0=C(B-b)/A$ is a dimensionless constant, $C_1$ is an integration constant, and  the equation $A/(B-b) = \Lambda_0 (1 + \gamma)/3\gamma$, following from (\ref{30}), is used. The integral in Eq. (\ref{32})  can be found for some specific values of $B$ and $b$, defined by the expressions (\ref{30}) through the EoS of matter  $w_m$ and the coupling constant $\gamma$. So, say for $ w_m = 0 \Rightarrow B = 3 $ and $ \gamma = 7 \Rightarrow b = 3/4$ in (\ref{32}), we have:
$$
a(t)=a_0\,\sinh^{2/3}\Big(\sqrt{\frac{8\Lambda_0}{21}}\,t\Big) \Rightarrow q(t) = -1 + \frac{3}{2}\cosh^{-2}\Big(\sqrt{\frac{8\Lambda_0}{21}}\,t\Big),
$$
where we put $C_1 =0$.  It follows that $q(0) = 1/2>0$ at the initial time, and then at some moment  $t_{0}$, defined by equation $q (t_{0})=0 $, the decelerated expansion followed by the accelerated one: $q(t>t_{0})<0$.

However, the model can be studied for the presence of acceleration in the general case with the help of definition (\ref{6}) and Eqs. (\ref{25}), (\ref{27}) and (\ref{29}). As a result, we can obtain  the following expression:
\begin{equation}\label{33}
q(x)= -1 + \frac{B}{2}-\frac{(B-b) e^{(B-b)x}}{2\Big[C_0 + e^{(B-b)x}\Big]},
\end{equation}
where $B\ne b \Leftrightarrow \gamma \ne \gamma_0$. As follows from this expression, the deceleration parameter evolves from $ q_i = q (- \infty) = -1 + B/2 $ at the beginning of expansion up to $ q_f = q (+ \infty) = -1 + b/2 $ to date, provided to $B> b$, but - from $q_f$ to $q_i$ for $B<b$. Note that the denominator in the formula (\ref{33}) is non-negative, as it follows from the right hand side of Eq. (\ref{29}) rewritten in the form  $ H^2 = (\Lambda_0 /3)(1 + \gamma^{-1}) \, e^{-Bx} [C_0 + e^{(B-b)x}] \geq 0 $. Since the latter must be true for all values of $ x \in (-\infty, +\infty) $, it also follows that $ C_0 \geq 0 $.  The evolution of $q(x)$  according to Eq. (\ref{33}) for several values of $ w_m $ and $ C_0 = 1 $ is shown in Fig. 5.

\section{Conclution}

~~~~Thus, we have studied the cosmological models in Lyra's geometry, supposing the so-called minimal interaction of matter both with the displacement vector field and with the dynamical $\Lambda$ - term. Besides general study, we have given some examples of exact
solution for the model under consideration.  Exact solutions of the dynamical  equations for our model in the cases $ w_m = -1 $ and $ w_m \ne-1 $, and under various assumptions about the evolution of the cosmological term are obtained. Surely, these solutions  can not
cover all possible applications of the present research.

Nevertheless, an interesting feature of these models with a quasi-vacuum EoS.  As can be observed in Fig. 2 and Fig. 4, these models begin to expand with a super-acceleration, and then asymptotically approach  a state of the constant acceleration or the non-accelerated state of expansion with $ q \geq 0 $, depending on the value of the coupling constant $ \gamma $.  It could be assumed that these models are relevant to the cosmological inflation.

Cosmological models, built in the last section with the assumption of proportionality between  the cosmological term and  the displacement field, demonstrate a tendency to shift from the slow expansion to the accelerated expansion.  That can be seen in Fig. 5. This behavior of these models suggests its relation to the phenomenon of the late-time acceleration, which is reliably confirmed by the observational data.

The interesting idea, which could be realized, consists of the combination of two cases considered above. Indeed, the behavior of the deceleration parameter in that case might reveal the inflationary  acceleration in the very beginning of expansion as well as the late-time acceleration.
Moreover, it is easy to obtain the exact solution for our model under the assumption of some different functions $\Lambda(t)$ widely discussed in the literature, namely $\Lambda \propto H^2(t)$, $\Lambda \propto \alpha H +\gamma H^2$ etc.
Further details and consequences of the model considered
here are in progress.

\end{document}